# Non-binary Unitary Error Bases and Quantum Codes

E. Knill

knill@lanl.gov, Mail Stop B265
Los Alamos National Laboratory
Los Alamos, NM 87545.

**Abstract**

Error operator bases for systems of any dimension are defined and natural generalizations of the bit/sign flip error basis for qubits are given. These bases allow generalizing the construction of quantum codes based on eigenspaces of Abelian groups. As a consequence, quantum codes can be constructed from linear codes over $\mathbb{Z}_n$ for any $n$. The generalization of the punctured code construction leads to many codes which permit transversal (i.e. fault tolerant) implementations of certain operations compatible with the error basis.
**Note:** This report is preliminary. Please contact the author if you wish to be notified of updates. A continuation can be found in "Group representations, error bases and quantum codes" [5].

## 1 Overview

Quantum error correction on products of qubits is increasingly well understood [10, 12, 3, 1, 6, 2]. See the references and the many papers on quantum computation and error correction on http://xxx.lanl.gov/ in the quant-ph section for background information on the subject. In this report, the problem of constructing quantum codes using systems of dimension greater than two is considered. Using the now traditional approach based on error operator bases pioneered by Steane [12], properties of good error operator bases for $n$-ary systems are discussed and examples of such bases are provided. These operator bases naturally generalize the bit/sign flip basis used for qubits. Many error bases allow generalizing the construction of Calderbank et al. [3] using eigenspaces of abelian groups. The general principle behind this construction is briefly discussed. It is shown that in principle any fidelity one error-correcting code can be obtained in this way. The generalized error bases lead naturally to codes based on $\mathbb{Z}_n$, the integers mod $n$. If Steane's construction [12] is used, then many interesting operations can be implemented transversally (i.e. fault tolerantly [11]). A basic set of such operations is described.

## 2 Error Operator Bases

The most commonly used basis of error operators on a two dimensional Hilbert space (a qubit) consists of the four matrices

$$I = \begin{pmatrix} 1 & 0 \\ 0 & 1 \end{pmatrix},$$
$$N = \begin{pmatrix} 0 & 1 \\ 1 & 0 \end{pmatrix},$$
$$S = \begin{pmatrix} 1 & 0 \\ 0 & -1 \end{pmatrix},$$
$$NS = \begin{pmatrix} 0 & -1 \\ 1 & 0 \end{pmatrix}.$$

The set $\mathcal{E}_2 = \{I, N, S, NS\}$ is an orthonormal basis of linear operators, where the inner product of operators $A$ and $B$ is given by $\mathrm{tr} A^\dagger B/2$. The set $\mathcal{E}_2$ and its tensor product extensions have many interesting algebraic properties. Those related to Clifford algebras are summarized in [2][1].

The goal of this report is to define error operator bases for dimensions greater than two and apply them to the construction of quantum error-correcting codes. Let $\mathcal{H}$ be a Hilbert space of dimension $n$. What properties should a good error operator basis $\mathcal{E}$ satisfy? First, the linear span of $\mathcal{E}$ should include all linear operators. Thus we require that $\mathcal{E}$ is a set of $n^2$ linearly independent operators. For the purpose of representing operators near the identity, we also want $I \in \mathcal{E}$. For being able to track error amplitudes reasonably well, it is desirable for $\mathcal{E}$ to be an orthonormal set of unitary operators. Thus we define a *unitary, orthonormal error operator basis* of $\mathcal{H}$ as a set $\mathcal{E} = \{E_1, \ldots, E_{n^2}\}$ of unitary operators on $\mathcal{H}$ such that $E_1 = I$, and $\operatorname{tr} E_i^\dagger E_j = n\delta_{i,j}$.

---

[1] Unfortunately a good review of the algebraic properties of $\mathcal{E}_2$ or of the relevant mathematical literature is not readily available.

Error operator bases can be used to track the evolution of a state in the presence of imperfect operations. If the intended state is $|\psi\rangle \in \mathcal{H}^{\otimes l}$, then the true state can be written in the form $\sum_i |e_i\rangle D_i |\psi\rangle$, where the $D_i$ are tensor products of error operators and the $|e_i\rangle$ are non-normalized states of the relevant environment. This is the error basis representation of the superoperator. The $|e_i\rangle$ need not be orthogonal. However, if the error operator basis is orthonormal, then $\sum_i \||e_i\rangle\|^2 = 1$. This follows from the fact that $\sum_{i,j} \langle e_i||e_j\rangle D_i^\dagger D_j = I$ and by taking traces on both sides. Another nice property of this representation is that the amplitude of the coefficient of the identity is the uniform fidelity amplitude introduced in [9]. See [7] for a discussion of the relationships between the error basis and the operator ensemble representations of superoperators.

For the remainder of this report, all error bases are orthonormal.

In order to properly track errors introduced by successive operations one has to compose error operators. Composing two members of an error basis yields a sum of operators according to an identity of the form

$$E_i E_j = \sum_k w_{ij,k} E_k,$$

where $w_{ij,k} = \operatorname{tr} E_k^\dagger E_i E_j / n$. Thus $\sum_k |w_{ij,k}|^2 = 1$. In general it is easier to track error amplitudes if only one error operator appears in the expression above. Thus a *nice* error basis satisfies

$$E_i E_j = w_{ij} E_{i*j}. \qquad (1)$$

for some operation $*$. By multiplying each $E_i$ by a suitable phase, it can be assumed that $\det E_i = 1$. A *very nice* error basis satisfies that each operator has determinant one. For example, $\mathcal{E}'_2 = \{I, iN, iS, NS\}$ is very nice for dimension two[2].

Some properties of nice error bases can easily be deduced.

**Theorem 2.1.** *If $\mathcal{E}$ is nice, then the operation $*$ induces a group on the indices with identity 1. The $w_{ij}$ have modulus 1. If it is very nice, then the $w_{ij}$ are n'th roots of unity.*

---
[2] $\mathcal{E}'_2$ generates a representation of the quaternions.

*Proof.* Because of orthonormality, an identity $E_i = \alpha E_j$ for a scalar $\alpha$ implies that $i = j$. Let $\bar{\mathcal{E}}$ be the group generated by $\mathcal{E}$. The subgroup $Z$ consisting of scalar multiples of the identity is normal. Each coset of $Z$ in $\bar{\mathcal{E}}$ contains exactly one member of $\mathcal{E}$, because two members of a coset differ only by a scalar multiple. Associating each member of the quotient $\bar{\mathcal{E}}/Z$ with the index $i$ of the unique $E_i$ contained in it establishes an isomorphism between $\bar{\mathcal{E}}/Z$ and the multiplicative structure on the indices.

The fact that $|w_{ij}| = 1$ follows from the fact that $|\det E_i| = 1$ and by taking determinants on both sides of Equation 1. The same process also shows that that $w_{ij}^n = 1$ when $\mathcal{E}$ is very nice. □

When the group is known, one can choose to index the error operators by the members of the group for convenience. For a very nice error basis, the elements of the form $e^{i2\pi/n}E_j$ form a group of order $n^3$. The subgroup generated by $\mathcal{E}$ is normal.

## 3   Examples of Nice Error Bases

A simple method for obtaining a nice error basis on $nm$ dimensions is to take the tensor products of the elements of error bases on $n$ and $m$ dimensions. Thus the "extra special group" described in [2] yields a nice error basis on $2^n$ dimensions.

Let $\omega$ be a primitive $n$'th root of unity. An error basis on $n$ dimensions not obtained by tensor products is generated by the two operators $D_\omega$ and $C_n$, where $(D_\omega)_{ij} = \delta_{i,j}\omega^i$ and $(C_n)_{ij} = \delta_{j,i+1 \mod n}$ (indices range between 0 and $n-1$). Thus $C_n$ is a cyclic permutation. The error basis is given by

$$\mathcal{E}_\omega = \{E_{i,j} = D_\omega^i C_n^j\}_{i,j\in\mathbb{Z}_n},$$

where we have chosen to index the error operators by elements of $G = \mathbb{Z}_n \times \mathbb{Z}_n$. The multiplication coefficients $\omega_{(i,j),(k,l)}$ are determined by the identity

$$C_n D_\omega = \omega D_\omega C_n.$$

Thus, the multiplication table is given by

$$E_{i,j} E_{k,l} = \omega^{jk} E_{i+k,j+l}.$$

This error basis is very nice if $n = 1 \mod 4$.

Error bases constructed by tensor products from the ones given above satisfy that the index group is abelian. Recently Sebastian Egner found a nice error basis with non-abelian index group by consulting character tables in the Neubüser catalog. The basis is obtained by choosing a representative from each coset of the center of the group generated by the following matrices:

$$A = \begin{pmatrix} 1 & 0 & 0 & 0 \\ 0 & -1 & 0 & 0 \\ 0 & 0 & 1 & 0 \\ 0 & 0 & 0 & -1 \end{pmatrix},$$

$$B = \begin{pmatrix} 0 & -i & 0 & 0 \\ 1 & 0 & 0 & 0 \\ 0 & 0 & 0 & i \\ 0 & 0 & 1 & 0 \end{pmatrix},$$

$$C = \begin{pmatrix} 0 & 0 & 1 & 0 \\ 0 & 0 & 0 & i \\ 1 & 0 & 0 & 0 \\ 0 & -i & 0 & 0 \end{pmatrix}.$$

The generators satisfy the following relations:

$$A^2 = C^2 = I, \ AB = -BA, \ AC = CA, \ BC = CB^{-1}, \ B^4 = -I.$$

The group generated by $A$, $B$ and $C$ has 32 elements. The center consists of $I$ and $-I$. The index group (which is given by the quotient over the center) is ismorphic to $\mathbb{Z}_2 \times D_4$. Methods for generating other such error bases and applications of group representations to the construction of error bases will be covered in a future report.

## 4 Error-correcting Codes

The purpose of a quantum error-correcting code is to restore a state which was initially in a coding subspace, after an interaction with an external system. For fidelity one error-correcting codes, the interactions that need to be corrected are given as a set of linear operators $\mathcal{A}$ acting on the supporting Hilbert space. In most of the work to date on fidelity one quantum error correcting codes, the supporting Hilbert space is a tensor product $\mathcal{Q}^{\otimes l}$ of qubits, and the operators in $\mathcal{A}$ are tensor products of error operators with a small number of non-identity factors. Thus it is natural to work with a tensor product error basis on $\mathcal{H}^{\otimes l}$.

A quantum code can be described as a subspace $\mathcal{C} \subseteq \mathcal{H}^{\otimes l}$. Sufficient and necessary conditions for the code to be $\mathcal{A}$-correcting have been given by several authors (see for example [1, 6]). The most commonly used condition requires that for a basis $\{|i_L\rangle\}_i$ of $\mathcal{C}$ and every $A, B \in \mathcal{A}$,

$$\langle i_L | A^\dagger B | j_L \rangle = \lambda_{A,B} \delta_{i,j}.$$

Let $P_\mathcal{C}$ be the projection on $\mathcal{C}$ and $I_\mathcal{C}$ the identity operator on $\mathcal{C}$. Then the condition can be restated in the form

$$P_\mathcal{C} A^\dagger B \propto I_\mathcal{C}.$$

The condition is satisfied if for each $X \in \mathcal{A}^\dagger \mathcal{A}$, $\mathcal{C}$ is either an eigenspace of $X$, or $X\mathcal{C}$ is orthogonal to $\mathcal{C}$. This is the basic idea underlying the construction described in [2]. There, $\mathcal{C}$ is obtained as an eigenspace of an abelian group $G$ of unitary operators such that for each $X \in \mathcal{A}^\dagger \mathcal{A}$, either $X \in G$, or $gX = \alpha Xg$ with $\alpha \neq 1$ for some element $g \in G$. The latter condition ensures that $X\mathcal{C}$ is one of the other eigenspaces of $G$, and hence orthogonal to the code.

It is a fact that every code which corrects $\mathcal{A}$ with fidelity one can be obtained by the construction above after a suitable change of representation of $\mathcal{A}$. This can be seen by using the syndrome characterization of codes given in [6]. If $I \in \mathcal{A}$, then $\mathcal{H} \cong \mathcal{S} \otimes \mathcal{C} \oplus \mathcal{R}$ with $A|0_S\rangle|\psi\rangle = \sum_i \alpha_{Ai}|i_S\rangle|\psi\rangle$, where the $|i_S\rangle$ form a suitable orthonormal basis of the syndrome space $\mathcal{S}$ and the code is identified with the subspace $|0_S\rangle \otimes \mathcal{C}$. Let $\mathcal{E}_\omega$ be the error basis on $\mathcal{S}$, with $\omega$ a $\dim \mathcal{S}$'th primitive root of unity. An operator is correctable if it is a linear combination of operators of the form $E_{i,j} \otimes I_\mathcal{C} \oplus R$ acting on the representation $\mathcal{H} \cong \mathcal{S} \otimes \mathcal{C} \oplus \mathcal{R}$. Here $R$ is any linear operator on $\mathcal{R}$. If operators of this form are used for the basic set $\mathcal{A}'$ of correctable operators, then it is the case that $A^\dagger B$ either has $|0_S\rangle \otimes \mathcal{C}$ as an eigenspace or $A^\dagger B|0_S\rangle \otimes \mathcal{C}$ is orthogonal to $|0_S\rangle \otimes \mathcal{C}$. We can even construct $|0_S\rangle$ as one of the eigenspaces of an abelian subgroup, namely the set of diagonal error operators extended to $\mathcal{R}$ by a suitable multiple of the identity. Although this does show the generality of the construction in [2], it is somewhat contrived unless we take advantage of additional structure.

# 5 Codes Based on Nice Error Bases

Let $\mathcal{E}$ be a nice error basis of $\mathcal{H}$ with abelian index group $G$. Let $\mathcal{E}^{\otimes l}$ be the corresponding tensor product error basis on $\mathcal{H}^{\otimes l}$. Elements of $\mathcal{E}^{\otimes l}$ can be described as sequences $\bar{D} = (D_1, D_2, \ldots, D_l)$, where $D_i \in \mathcal{E}$ acts on the $i$'th factor of $\mathcal{H}^{\otimes l}$. The index $\bar{d} = (d_1, \ldots, d_l)$ of $\bar{D}$ is the sequence of indices of the $D_i$ in $\mathcal{E}$. Thus $\bar{d}$ is in the $l$-fold product of the index group of $\mathcal{E}$. The weight $|\bar{d}|$ of $\bar{d}$ is given by the number of non-identity elements in $\bar{D}$. Define $\bar{d} \cdot \bar{d}' = \prod_i \omega_{d_i d'_i} \bar{\omega}_{d'_i d_i}$. This product gives the commutation relationships between the elements of $\mathcal{E}^{\otimes l}$. That is

$$\bar{D}\bar{D}' = \bar{d} \cdot \bar{d}' \bar{D}' \bar{D} \,.$$

Define $\bar{d}$ to be orthogonal to $\bar{d}'$ if $\bar{d} \cdot \bar{d}' = 1$, i.e. if $\bar{D}$ commutes with $\bar{D}'$.

Let $\mathcal{A}$ be a subset of $\mathcal{E}^{\otimes l}$. Let $H$ be an abelian subgroup of $\bar{\mathcal{E}}^{\otimes l}$ such that $\mathcal{A}^\dagger \mathcal{A} \setminus \mathbb{C}H$ does not contain a member orthogonal to all elements of $H$. Then any eigenspace of $H$ is a code which can correct $\mathcal{A}$.

As a useful example, consider $\mathcal{E} = \mathcal{E}_\omega$ with $\omega$ an $n$'th primitive root of unity. In this case the index group is $\mathbb{Z}_n \times \mathbb{Z}_n$, so the indices of elements of $\mathcal{E}_\omega^{\otimes l}$ can be written as pairs $(x, y)$ of vectors over $\mathbb{Z}_n$, where the first component refers to the diagonal operators and the second to the cyclic permutations of $\mathcal{E}_\omega$. The product of indices is defined by $(x, y) \cdot (x', y') = \omega^{y \cdot x' - x \cdot y'}$, where the products in the exponent of $\omega$ are the usual inner product of vectors over $\mathbb{Z}_n$.

Let $C$ and $D$ be codes of length $l+1$ over $\mathbb{Z}_n$ with $D$ the dual of $C$ and vice versa. Assume that every member of $\mathbb{Z}_n$ occurs as a last element of code words of $C$. Let $C'$ be obtained from $C$ by puncturing at the last element. Let $C'_0$ be the subcode of $C'$ obtained from words of $C$ with last element 0. Define $D'$ and $D'_0$ similarly. The dual of $C'$ is $D'_0$ and the dual of $C'_0$ is $D'$. Let $e_1$ be a word in $C$ with last element 1 and define $e_i = ie_1$. Let $e'_i$ be the word obtained by removing the last element of $e_i$. Assume that $e'_i \notin C'_0$ for $i \neq 0$. The group H generated by the set of elements of $\mathcal{E}_\omega^{\otimes l}$ with indices $(x, y)$, $x \in C'_0$, $y \in D'_0$ is abelian and an eigenspace $\mathcal{C}$ is given by linear combinations of

$$|i_L\rangle = \sum_{x \in C'_0 + e'_i} |x\rangle \,.$$

Here we used the usual basis of $\mathcal{H}$ and $\mathcal{H}^{\otimes l}$ labeled by code words and consistent with the representation of the matrices in $\mathcal{E}$.

If both $C'$ and $D'$ have minimum weight at least $2e+1$, then any error pattern of weight at most $e$ can be corrected. This is true because if $(x', y')$ is orthogonal to all elements of the group $H$, then $x' \in C'$ and $y' \in D'$. For error operators $A$ and $B$ of weight at most $e$, $A^\dagger B$ has weight at most $2e$.

The actual error correcting properties of $\mathcal{C}$ are determined to some extent by an associated syndrome decoding method. Let $\{c_x\}_x$ and $\{d_y\}_y$ be complete sets of coset representatives for the cosets of $C'$ and $D'$, respectively. Assume that the indices are elements of the respective quotients. Thus $\{c_x + C'\}$ and $\{d_y + D'\}$ partition $\mathbb{Z}_n^l$. Let $E(c_x, d_y)$ be the error operator with index $(c_x, d_y)$. Let

$$\begin{aligned}|c_x, d_y\rangle|i_L\rangle &= E(c_x, d_y)|i_L\rangle \\ &= \sum_{z \in C_0'} \omega^{c_x \cdot z} |z + e_i' + d_y\rangle.\end{aligned}$$

This induces a natural representation of $\mathcal{H}^{\otimes l} \cong \mathcal{S} \otimes \mathcal{C}$. It also determines a recovery procedure: Measure the syndrome and return it to $(c_0, d_0)$ by applying $E(c_x, d_y)^\dagger$. The representatives of the cosets chosen determine which errors are corrected by the procedure. Note that the effect of an uncorrectable error is to induce a unitary operation (equivalent to an error in the error basis) on the code after correction. These operations can be exploited for fault tolerant implementation of gates.

The syndrome measurement can be performed as described in [12] by measuring classical syndromes in two bases. This can be accomplished fault tolerantly [11] if so desired, by using a suitable generalization of "cat states" and their preparation. To see that the two bases approach works, it suffices to observe that if $F$ is the Fourier transform matrix (based on $\omega$, $F_{ij} = \omega^{ij}$), then $C_n = F D_\omega F^\dagger$. Thus every error of type $D_\omega^k$ looks like a cyclic permutation in the Fourier transformed basis.

## 6   Transversally Implementable Operations

One of the useful properties of punctured codes for quantum error correction is that many operations on the encoded state can be implemented by acting directly and independently on qubits without decoding. Such transversally implemented operations allow for at least some computations to be performed fault tolerantly [11, 8].

Consider the punctured coding construction of the previous section. Some linear operations are particularly easy to implement directly. For example to add 1 mod $n$ to each code word, it suffices to apply the cyclic permutation $C_n^{(e_1')_i}$ to the $i$'th underlying system. In effect this applies the cyclic permutation $C_n$ to the code. One can also apply $D_\omega$ to the code, for example by changing to the Fourier transformed basis and using the same technique as for $C_n$. Of course, the transformation can be avoided by applying suitable powers of $D_\omega$ directly. This permits transversal implementation of the error group on the encoded state. It is interesting to observe that measurement in both the $|i_L\rangle$ and the Fourier transformed basis is possible. If $C$ is self dual, then the Fourier transform can be applied to the code by applying it to each system. Finally, it is possible to instantiate the generalization of the controlled-not operation. The operation defined by $N|i_L\rangle|j_L\rangle = |i_L\rangle|(j+i)_L\rangle$ is obtained by applying it to each corresponding system of the two encoded states.

The operations introduced so far are in the normalizer of the error group and hence do not generate a sufficiently dense set for the purpose of quantum computation. Methods involving state preparation and measurement such as those used in [11, 8] are required to complete the set.

# 7 Acknowledgements

Thanks to Thomas Beth and Markus Grassl for valuable discussions and to Sebastian Egner for providing the first example of a nice error basis with nonabelian index group.